\documentclass[aps,prd,preprint,onecolumn,nofootinbib]{revtex4-2}

\usepackage{amsmath,amsfonts,amssymb}
\usepackage{graphicx}
\usepackage{color}
\usepackage{hyperref}
\usepackage{bm}
\usepackage{mathrsfs}

\usepackage{orcidlink}

\def\be{\begin{eqnarray}}
\def\ee{\end{eqnarray}}
\def\nn{\nonumber}

\begin{document}

\title{
Causal structure of black holes immersed in a Chaplygin-like dark fluid environment: Horizons and singularities
}

\author{Rodrigo Dal Bosco Fontana\,\orcidlink{0000-0002-8835-7447}}
\email[Corresponding author: ]{rodrigo.dalbosco@ufrgs.br}
\affiliation{Federal University of Rio Grande do Sul - UFRGS, Campus Tramandaí-RS, Estrada Tramandaí-Osório, Emboaba, Tramandaí 95590-000, Brazil}

\author{Jeferson de Oliveira\,\orcidlink{0000-0002-0874-3613}}
\email{jeferson.oliveira@ufmt.br}
\affiliation{Instituto de Física, \\ Universidade Federal de Mato Grosso, Cuiabá, 78060-900, MT, Brazil}

\date{\today}

\begin{abstract}
In the present work, we study the causal structure of spherically symmetric black holes immersed in a Chaplygin-like dark fluid, emphasizing the impact of the fluid parameters on curvature and horizon formation. We show that the spacetime curvature is significantly stronger than in its similar counterpart, the Reissner–Nordström–de Sitter geometry with the same mass and charge, leading to modifications of the internal causal structure. For the presence of horizons the Chaplygin black hole possesses an upper bound $Q\approx0.556219 M$, which is much smaller than that for Reissner-Nordstr\"om spacetime $Q_{critical}=M$ or of the Reissner-Nordstr\"om-de Sitter case $Q_{critical}=3M/2\sqrt{2}$, indicating that the black holes immersed in a Chaplygin-like dark fluid reach the extremal regime more easily. We derive a second critical condition for the Chaplygin cosmological parameter $B$, $B_c Q_c = 4/3^9$, setting an upper bound on $B$ for a multi-horizon solution.

\end{abstract}

\maketitle

\section{Introduction}

Black holes represent the most extreme solutions of general relativity, embodying the nonlinear regime of gravity where spacetime curvature, causal structure, and gravitational dynamics are driven to their theoretical limits. From a theoretical perspective, they provide a framework in which strong field effects can be explored, allowing fundamental notions such as event horizons, spacetime singularities, and global causality to be rigorously defined and tested \cite{Chandrasekhar1983} \cite{HawkingEllis1973} \cite{Tipler:1978zz} \cite{wald:1984}. In recent years, this  theoretical domain has been transformed by observations. The direct detection of gravitational waves from compact binary mergers by the LIGO-VIRGO-KAGRA (LVK) collaboration has confirmed the existence of black holes as dynamical astrophysical objects and enabled quantitative tests of gravity in the highly nonlinear regime \cite{Abbott2016GW150914} \cite{Abbott2016TestsGR} \cite{Berti2015TestingGR} \cite{Berti2015TestingGR} \cite{Maggiore1,Maggiore2} \cite{GW250114_PRL_2025}. In parallel,  images of supermassive black holes obtained by the Event Horizon Telescope (EHT) have provided direct observational access to the photon sphere region and to the immediate vicinity of the event horizon \cite{EHT2019M87I,EHT2019M87II,EHT2022SgrAI,EHT2022SgrAV,EventHorizonTelescope:2024hpu}. Together, these observations have established black holes as precision probes of strong gravity.

These developments motivate the study of black holes beyond the idealized vacuum regime. Astrophysical black holes are embedded in environments containing matter and fields, whose presence can modify both local and global properties of the spacetime \cite{Cardoso2022LivingReviews,  Bonanno:2020fgp, ref1b, ref2b, ref3b}. In particular, black holes in the presence of nonstandard or exotic fluids allow one to investigate how matter content influences causal structure, horizon behavior, and the nature of spacetime singularities, as well as their imprints in gravitational wave signals an quasinormal modes spectrum \cite{Visser2004DirtyBH, Cuadros-Melgar:2020shz, Cuadros-Melgar:2021sjy,deOliveira2025CQG, qnmbk, us3,  ref4b}. 

\textcolor{black}{In this context, models in which black holes are embedded in dark fluids provide a useful framework to investigate environmental effects on black hole physics. Among the proposed candidates, the Chaplygin gas and its generalizations have attracted considerable attention as unified dark sector models capable of describing both dark matter and dark energy within a single fluid. In particular, the Chaplygin-like dark fluid considered is characterized by the equation of state $p=-B/\rho$, which leads to an energy density that interpolates between a matter-like behavior at small radial distances and a cosmological constant at large scales. More specifically, for small values of the radial coordinate the energy density behaves as $\rho \sim q/r^3$, while at large distances it approaches a constant value $\rho \sim \sqrt{B}$, effectively acting as a positive cosmological constant \cite{Li2023ChaplyginPRD}\cite{Li2024ChaplyginJCAP048}.}

\textcolor{black}{
This property makes black hole solutions immersed in Chaplygin-like dark fluids particularly interesting for astrophysical applications, since they allow one to simultaneously model local matter effects near the event horizon and the large-scale cosmological expansion. Moreover, previous studies have shown that the presence of the Chaplygin-like fluid can modify observable properties of black holes, including their optical appearance and shadow, which may be constrained by Event Horizon Telescope observations \cite{Zahid2024CQGChaplyginShadow}. In addition, the interaction between the surrounding fluid and perturbative fields can affect the quasinormal mode spectrum \cite{Becar:2024agj}, potentially leaving imprints on the gravitational-wave ringdown signal. Therefore, investigating perturbations in these backgrounds provides a useful way to explore how environmental effects associated with dark sector models may influence observable signatures of black holes.}

Black holes immersed in Chaplygin-like  dark fluids arise as exact, spherically symmetric solutions of the Einstein equations sourced by an effective fluid characterized by a nonlinear equation of state \cite{Li2024ChaplyginJCAP048,Becar:2024agj,Li2023ChaplyginPRD,Zahid2024CQGChaplyginShadow}. Such a fluid interpolates between a matter behavior at small radial distance and a positive cosmological constant component at large distances, providing a description of a medium that captures both local and asymptotic matter effects.

In this work, we analyze the causal structure of spherically symmetric black hole solutions immersed in a Chaplygin-like dark fluid. We examine the global properties of the spacetime by identifying horizons and singularities and by determining how their existence and multiplicity depend on the parameters that characterize the solution. The analysis allow us to establish bounds on the parameters $Q$ and $B$ delimiting the regions of parameter space where the spacetime admits an event horizon, a Cauchy horizon and a cosmological horizon, as well as configurations corresponding to extremal black holes. In Sec. \ref{sec2}, we review the black hole solution, summarizing its main properties. In Sec. \ref{sec3}, we investigate how the causal structure of the spacetime depends on the parameters that characterize the solution, with particular emphasis on the behavior of horizons and singularities. Finally, in Sec. \ref{sec4}, we present a discussion of the results and our final remarks.


\section{Background black hole solution} \label{sec2}

In this section, we review the main properties of the spherically symmetric black hole solution immersed in a Chaplygin-like dark fluid, which will serve as the background spacetime throughout this work. \textcolor{black}{Following the procedure adopted in previous studies of black holes immersed in Chaplygin-like dark fluids \cite{Li2024ChaplyginJCAP048, Zahid2024CQGChaplyginShadow, Becar:2024agj}, we briefly summarize the main steps leading to the relevant expressions used in this work.}

We consider a static and spherically symmetric black hole solution sourced by a Chaplygin-like dark fluid, described by the line element
\begin{equation}
\label{le1}
ds^{2} = -f\,dt^{2} + f^{-1}dr^{2} + r^{2}\left(d\theta^{2} + \sin^{2}\theta\,d\phi^{2}\right),
\end{equation}
where the lapse function $f \equiv f(r)$ is determined by the Einstein equations coupled to an anisotropic energy--momentum tensor,
\begin{equation}\label{tensorEM}
T_{\mu\nu} = \rho\, u_\mu u_\nu + p_r\, k_\mu k_\nu + p_t\, \Pi_{\mu\nu}.
\end{equation}
Here, $\rho$ denotes the energy density of the fluid, while $p_r$ and $p_t$ are the radial and tangential pressure components, respectively. The vector $u^\mu$ is the four--velocity of the fluid, and $\Pi_{\mu\nu} = g_{\mu\nu} + u_\mu u_\nu - k_\mu k_\nu$ is a projection tensor, where $k^\mu$ is a spacelike unit vector orthogonal to $u^\mu$, satisfying
\begin{equation}
u_\mu u^\mu = -1, \qquad k_\mu k^\mu = 1, \qquad u_\mu k^\mu = 0 .
\end{equation}
The condition $p_r = -\rho$ is imposed to ensure a static configuration of the fluid and to guarantee the continuity of the energy density across the event horizon \cite{Li2024ChaplyginJCAP048}. Assuming a nonlinear equation of state of Chaplygin type,
\begin{equation}
p = -\frac{b}{\rho},
\end{equation}
with $b>0$, the tangential pressure is given by $p_t = \tfrac{1}{2}\rho - \tfrac{3b}{2\rho}$. \textcolor{black}{Such expression for the tangential pressure $p_t$ follows from the anisotropic stress–energy tensor of the Chaplygin-like dark fluid (\ref{tensorEM}) together with the equation of state $p=-B/\rho$, the condition $p_r=-\rho$ \cite{Li2023ChaplyginPRD}.}

From the energy-momentum tensor~\eqref{tensorEM}, one obtains the exact expression for the energy density,
\begin{equation}\label{densidade_energia}
\rho(r) = \sqrt{b + \frac{q^{2}}{r^{6}}},
\end{equation}
where $q$ is a positive constant related to the intensity of the Chaplygin-like dark fluid.
\textcolor{black}{This expression for the energy density is obtained by solving the Einstein equations for a static and spherically symmetric spacetime (\ref{le1}) sourced by the Chaplygin-like dark fluid (\ref{tensorEM}):
\begin{eqnarray}
    \frac{1}{r^2}\left(f+rf'-1\right) &=& -\rho\\
    \frac{1}{2r}\left(2f' + rf''\right) &=& \frac{1}{2}\rho -\frac{3b}{2\rho}
\end{eqnarray}
}
As discussed in \cite{Becar:2024agj}, at small radial distances the energy density behaves as
\begin{equation}
\rho(r) \approx \frac{q}{r^{3}},
\end{equation}
which corresponds to a matter profile with $\rho \propto r^{-3}$. In the asymptotic regime, the energy density approaches a constant value,
\begin{equation}
\rho(r) \approx \sqrt{b},
\end{equation}
effectively mimicking a positive cosmological constant at large distances.

\textcolor{black}{It is also worth noting that the energy density remains finite at the event horizon $r=r_h$. The regularity of the fluid across the event horizon is given by the condition $p_r=-\rho$. As discussed in \cite{Li2023ChaplyginPRD}, when crossing the event horizon the roles of the temporal and radial coordinates interchange, since inside the horizon $g_{tt}>0$ and $g_{rr}<0$, implying that the coordinate $r$ becomes timelike. In this situation the energy density and pressure remain continuous across the horizon only if the condition $p_r=-\rho$ is satisfied. Otherwise, if $p_r\neq-\rho$ and $\rho(r_h)\neq0$, the pressure would become discontinuous at the horizon and the corresponding solution would necessarily be dynamical. Therefore, imposing $p_r=-\rho$ guarantees that the Chaplygin-like dark fluid remains static and that the stress–energy tensor is continuous across the horizon.}

Solving the Einstein equations with (\ref{densidade_energia}) and (\ref{tensorEM}), we obtain the lapse function
\begin{equation}
f(r) = 1-\frac{2M}{r}
-\frac{r^{2}}{3}\sqrt{b+\frac{q^{2}}{r^{6}}}
+\frac{q}{3r}\operatorname{arcsinh}\!\left(\frac{q}{\sqrt{b}\,r^{3}}\right),
\end{equation}
where the parameters $b$ and $q$ characterize the contribution of the Chaplygin-like dark fluid to the spacetime geometry. \textcolor{black}{The metric function $f(r)$ is obtained by substituting the energy density profile of (\ref{densidade_energia}) into the Einstein equations. Integrating the resulting differential equations yields the analytical form of the lapse function shown above.}

Here we reparameterize the constants $b=9B$ and $q=3Q$ in order to simplify the lapse function,
\begin{equation}
\label{lapse1a}
f(r) = 1-\frac{2M}{r}
-r^{2}\sqrt{B+\frac{Q^{2}}{r^{6}}}
+\frac{Q}{r}\operatorname{arcsinh}\!\left(\frac{Q}{\sqrt{B}\,r^{3}}\right).
\end{equation}
In the limit $Q \rightarrow 0$, the solution reduces to the Schwarzschild--de Sitter black hole, with an effective cosmological constant $\Lambda_{\rm eff} = 3\sqrt{B}$. In the limit where both parameters vanish, $Q \rightarrow 0$ and $B \rightarrow 0$ the metric consistently reduces to the Schwarzschild solution.

At large radial distances, the lapse function asymptotically behaves as
\begin{equation}\label{lapse}
f(r) \approx 1 - \sqrt{B}\, r^{2},
\end{equation}
indicating the presence of a pseudo-cosmological horizon. As discussed in \cite{Li2024ChaplyginJCAP048}, the parameter $q$ primarily controls the existence and location of the event horizon $r_{+}$, while $B$ determines the position of the pseudo-cosmological horizon $r_c$. 

\begin{figure}[htb]
\includegraphics[width=9.8 cm]{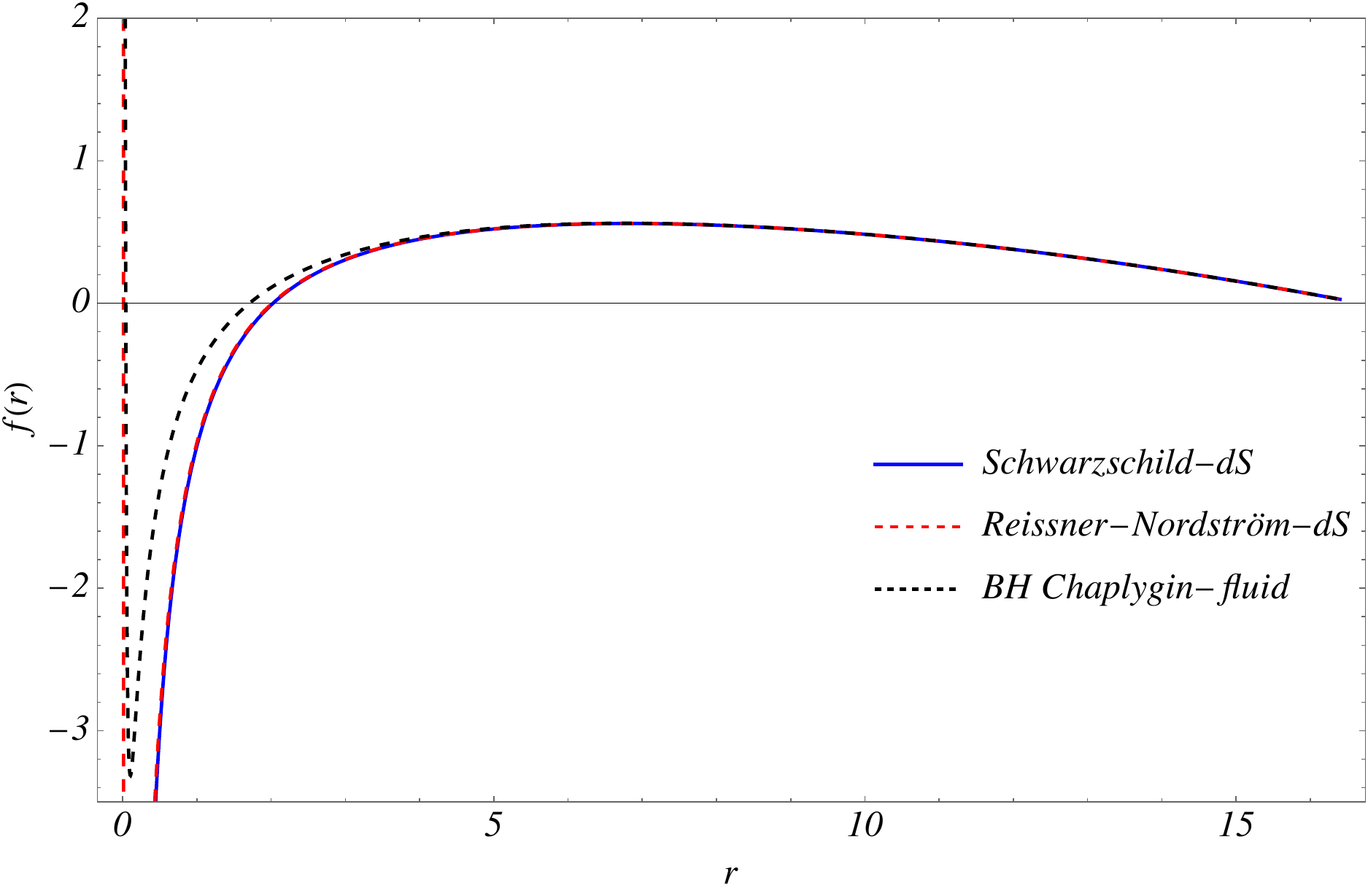}
\caption{\textcolor{black}{Behavior of the lapse function $f(r)$ for three different spacetimes: Schwarzschild–de Sitter ($r_h=2.02$), Reissner–Nordström–de Sitter ($r_h=2.01$), and a black hole immersed in a Chaplygin-like dark fluid ($r_h=1.75$). The parameters used are $M=1$, $Q=0.15$, and $B=10^{-5}$. }}
\label{fig1_1}
\end{figure}

The behavior of the lapse function (\ref{lapse1a}) is illustrated in Fig. \ref{fig1_1}, where we compare the present solution with the Schwarzschild–de Sitter and Reissner–Nordström–de Sitter cases, highlighting the modifications introduced by the Chaplygin-like dark fluid.

\section{Curvature scalars, parameter space and horizon structure}\label{sec3}
The determination of the causal structure of the black hole solution we are characterizing starts with the Ricci and Riemmann curvature invariants. In terms of (\ref{le1}) we have
\be
\label{c3}
\mathcal{R} & = & -\frac{r^2f'' + 4rf'+2f-2}{r^2}, \\
\label{c4}
{\mathcal{R}}_{\mu \nu} {\mathcal{R}}^{\mu \nu} & = & \frac{4rf'(r^2f''-2) r^4{f''}^2 + 8f(rf'-1) + 4f^2 +4}{2r^4}, \\
\label{c5}
{\mathcal{R}}_{\mu \nu \alpha \beta} {\mathcal{R}}^{\mu \nu \alpha \beta} & = & \frac{4((f-1)^2 + r^2{f'}^2)}{r^4} + {f''}^2,
\ee
in which prime denotes derivative relative to $r$. Scalars (\ref{c3}) to (\ref{c5}) when applied to the lapse function (\ref{lapse}) yield
\be
\label{c6}
\mathcal{R} & = & \frac{3 \left(4 B^2 r^{12}+5 B Q^2 r^6+Q^4\right)}{r^3 \left(B r^6+Q^2\right)^{3/2}}, \\
\label{c7}
{\mathcal{R}}_{\mu \nu} {\mathcal{R}}^{\mu \nu} & = & \frac{9}{2} \left(B \left(8-\frac{9 Q^2}{B r^6+Q^2}\right)+\frac{5 Q^2}{r^6}\right), \\
\nn
{\mathcal{R}}_{\mu \nu \alpha \beta} {\mathcal{R}}^{\mu \nu \alpha \beta} & = & \frac{24 M \left(2 M-\frac{3 Q^2}{\sqrt{B r^6+Q^2}}\right)+\frac{81 Q^4}{B r^6+Q^2}+24 B r^6-12 Q^2}{r^6} \\
\label{c8}
& & \hspace{1.0cm} +\frac{12 Q \sinh ^{-1}\left(\frac{Q}{\sqrt{B} r^3}\right) \left(\frac{3 Q^2}{\sqrt{B r^6+Q^2}}+Q \sinh ^{-1}\left(\frac{Q}{\sqrt{B}
   r^3}\right)-4 M\right)}{r^6},
\ee
The curvature invariants have a physical singularity at $r=0$ if $Q \neq 0$ or $M \neq 0$ represented as a timelike or spacelike structure depending on the presence of $M$ and $Q$. The curvature \textcolor{black}{near the singularity} is stronger then in the counterpart solution, the Reissner-Nordström-de Sitter spacetime: in this case the Kretschman scalar diverges as $r^{-6}$ at $r=0$, and for the Chaplygin black hole the Arcsinh function enhances the singularity character at $r=0$.

A central point in the causal structure of the black hole, besides the  singularity nature is the existence of horizons. Considering the \textcolor{black}{diagonal line-element (\ref{le1}), Killing horizons are defined by the condition $g_{tt}=0$. Those surfaces represent also event horizons in the static spherically symmetric geometries where $g_{tt}g_{rr}=-1$. Thus, for our Chaplygin black hole, horizons are defined by the singular points of (\ref{lapse1a})
\be
\label{eea}
f=0 
\ee
The function $g_{tt}$ also establishes the direction of curves in light-cone generators thus completely enlightening the causal structure of the black hole. It provides the usual interpretation of the coordinates $r$ and $t$ in each block of a Penrose diagram, similarly to the counterpart solution of RNdS. We present such a diagram at the end of this section, after analyzing the conditions under which horizons exist.}

We begin the analysis of $f$ by examining its behavior in the manifold asymptotic regions,
\be
\label{as1}
&& \lim_{r \to 0 } f \rightarrow \infty , Q\neq 0 \hspace{1.5cm} \lim_{r \to 0 } f \rightarrow -\infty , Q = 0 \hspace{0.3cm}  \& \hspace{0.3cm}  M \neq 0 \\
\label{as2}
&& \lim_{r \to \infty } f \rightarrow -\infty , B \neq 0 \hspace{1.2cm} \lim_{r \to \infty } f \rightarrow 1 , B = 0 
\ee
Relations (\ref{as1}-\ref{as2}) establish the existence of at last one horizon whenever the Chaplygin fluid is present with $Q$ and $B$. In specific cases we discussed in the previous section the metric reduces to well-known solutions of general relativity (Schwarzschild and Schwarzschild-de Sitter). Considering the existence of at least one singular point for $f$, we want to establish the maximum number of roots for $f=0$ and, particularly under what conditions those solutions exist. With that purpose we propose two reparametrizations of $r$ in what follows.

Firstly, for the sake of simplicity, we can eliminate $M$ of $f$ by rewriting the radial coordinate and spacetime constants,
\be
\label{c9}
r & \equiv & 2M x, \\
Q & \equiv & 2M \mathcal{Q}, \\
B & \equiv & \frac{\mathcal{B}}{16M^4},
\ee
bringing 
\be
\label{lp2}
f(x) = 1-\frac{1}{x}
- \sqrt{\mathcal{B}x^4+\frac{\mathcal{Q}^{2}}{x^{2}}}
+\frac{\mathcal{Q}}{x}\operatorname{arcsinh}\!\left(\frac{\mathcal{Q}}{\sqrt{\mathcal{B}}\,x^{3}}\right),
\ee
Considering the existence of nonlinear terms in (\ref{lp2}) we can not directly determine its singular points. Even though, the analysis can be carry further with the derivative of $f$,
\be
\label{df}
\frac{\partial f}{\partial x} = \frac{1}{x^2}\left( 1 - 2\sqrt{\mathcal{Q}^2 + \mathcal{B}x^6} - \mathcal{Q}\operatorname{arcsinh}\left(\frac{\mathcal{Q}}{\sqrt{\mathcal{B}}x^3}\right) \right),
\ee
and a new radial coordinate $y=\frac{\mathcal{Q}}{\sqrt{\mathcal{B}}x^3}$ that brings both the function and its derivative to a suitable form,
\be
\label{lap3}
f(y) & = & 1 - \left( \mathcal{Q}^2 \sqrt{\mathcal{B}} y \right)^{1/3} \left(\frac{1}{\mathcal{Q}} + \sqrt{1+\frac{1}{y^2}} -  \operatorname{arcsinh} y \right) \\
\label{df2}
\dot{f} \equiv \frac{\partial f}{\partial y} & = & \left( \frac{\mathcal{Q}^2 \sqrt{\mathcal{B}}}{ 3y^2} \right)^{1/3} \left(-\frac{1}{\mathcal{Q}} + 2\sqrt{1+\frac{1}{y^2}} +  \operatorname{arcsinh} y \right) \equiv \left( \frac{\mathcal{Q}^2 \sqrt{\mathcal{B}}}{ 3y^2} \right)^{1/3} g(y).
\ee

\begin{figure}[htb!]
\includegraphics[width=10 cm]{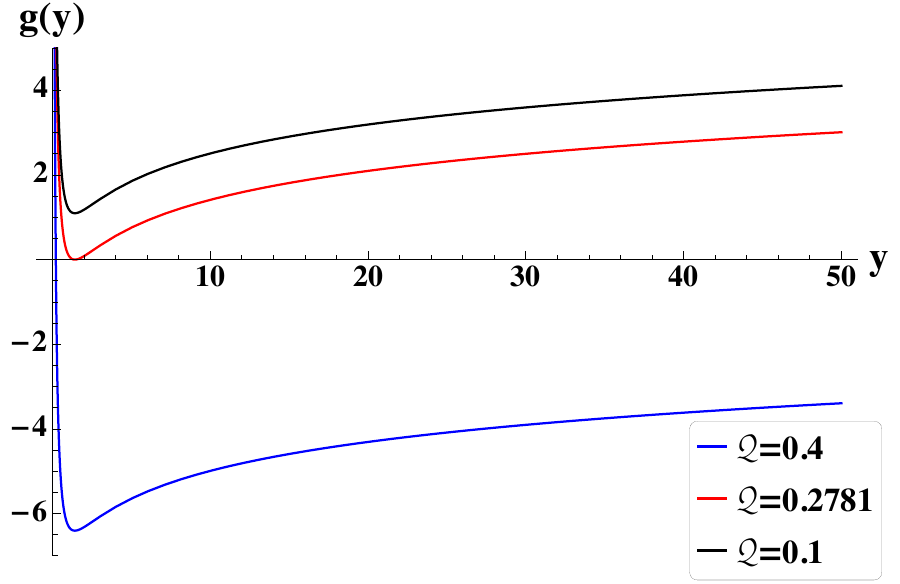}
\caption{Profile of the function $g(y)$ for different charges.}
\label{fig1}
\end{figure}
\begin{figure}[htb!]
\includegraphics[width=10 cm]{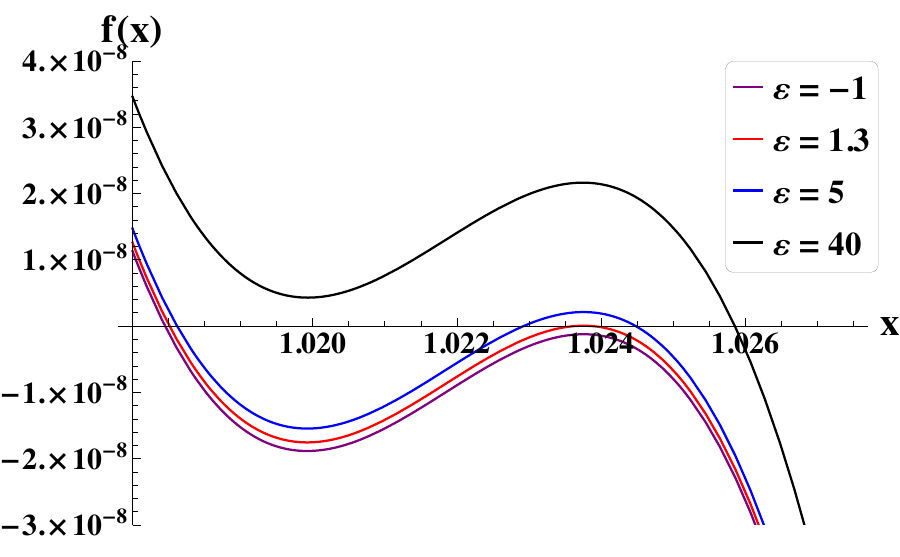}
\caption{The lapse function of Chaplygin black holes  in the near extremal regime in diomensionless coordinate $x$ . The geometry parameters read , $\mathcal{B}=0.00212314$ and $\mathcal{Q}=0.278107533283 + 5*10^{-10}\varepsilon$. The critical values for charge and cosmological term are $2M\mathcal{Q} =Q \simeq 0.55621906656636756M$ and $16M^4\mathcal{B} = B \simeq 0.00212317216M^4$.}
\label{fig2}
\end{figure}

   
\unskip
The interesting feature about $g(y)$ is that its profile closely resembles the Newtonian potential of the two-body problem, exhibiting exactly one saddle point for $y \geq 0$.  We display some exemplary plots of $g$ in figure \ref{fig1}. Considering the derivative of $g$, 
\be
\label{ddf}
\frac{\partial g}{\partial y} = \frac{y^2-2}{y^2\sqrt{y^2+1}},
\ee
we have just one solution for $\dot{g}=0$ in the region $y \geq 0$, $y=\sqrt{2}$ $(\equiv y_s )$. In such case $\dot{f}\Big|_{y_s}$ is a minimum saddle point. As a consequence, if $\dot{f} \Big|_{y_s} > 0 $ the spacetime has precisely one horizon since $f$ has no stationary points. That condition defines the threshold charge of the black hole as
\be
\dot{f} \Big|_{y=\sqrt{2}} = 0, 
\ee
setting the upper bound of $\mathcal{Q}$,
\be
\label{crit1}
\mathcal{Q}_{th} = \left(  2\sqrt{1+\frac{1}{\sqrt{2} ^2}} +  \operatorname{arcsinh} \sqrt{2}  \right)^{-1} \simeq 0.278109533283,
\ee
aproximately
\be
\label{crit1a}
Q_{critical} \sim 0.556219 M.
\ee
We display a panel with the near extremal black hole considering the critical value of charge (and of cosmological term) in figure \ref{fig2}. Interesting enough the threshold for the Chaplygin solution is much smaller than that of a Reissner-Nordström spacetime $Q_{critical} = M$ or of a Reissner-Nordström-de Sitter black hole $Q_{critical} = 3M/2\sqrt{2}$ \cite{rodjefala}.

\unskip
The existence of two stationary points in $f$ settles the first condition for a three horizon solution. In the threshold situation, one of those points must still coincide with a singular point of $f$. That can be written with (\ref{lap3}) and (\ref{df2}) as
\be
\label{cd1}
f\Big|_{y_t} = 0, \\
\label{cd2}
\dot{f}\Big|_{y_t} = 0.
\ee
Equation (\ref{cd2}) can be also expressed as $g(y_t)=0$ and
\be
\label{cd3}
\operatorname{arcsinh} y_t = \frac{1}{\mathcal{Q}} - 2 \sqrt{1+\frac{1}{y_t^2}}.
\ee
Substituting $y_t = \mathcal{Q}\mathcal{B}^{-1/2}x_t^{-3}$ into (\ref{cd3}) and then the arcsinh term into (\ref{cd1}) we obtain
\be
\label{pol1}
H(z) \equiv \mathcal{B}z^3 - \frac{1}{9}z + \mathcal{Q}^2 = 0
\label{cd4}
\ee
in which $z \equiv x_t^2$. $H(z)$ is a third order polynomial possessing three roots $z_i$ expressed as
\be
\label{z1}
z_1 & = & \frac{1}{3}\left(\left( \frac{2}{9 } \right)^{1/3}\delta^{-1/3} + \frac{1}{6^{1/3} \mathcal{B}}\delta^{1/3} \right) \\
\label{z2}
z_2 & = & \frac{-1+i\sqrt{3}}{972^{1/3}} \delta^{-1/3} + \frac{(-1-i\sqrt{3})}{6^{4/3}\mathcal{B}}\delta^{1/3} \\ 
\label{z3}
z_{\textcolor{black}{3}} & = & \frac{-1-i\sqrt{3}}{972^{1/3}} \delta^{-1/3} + \frac{(-1+i\sqrt{3})}{6^{4/3}\mathcal{B}}\delta^{1/3} 
\ee
with 
\be
\label{delta}
\delta = -3^4 \mathcal{B}^2\mathcal{Q}^2 + \sqrt{(3^4\mathcal{B}^2 \mathcal{Q}^2 )^2 - \frac{4\mathcal{B}^3}{3}}.
\ee
In the above expression, depending on the values of $\mathcal{B}$ and $\mathcal{Q}$ we may have either $\delta \in \mathbb{R}$ or $\delta \in \mathbb{C}$.  \textcolor{black}{In the first case, considering the expression (\ref{delta}) we see that $\delta < 0$ bringing as a consequence $z_1 < 0$ and $z_2,z_3 \in \mathbb{C}$ forbidding positive real solutions for $x_t$ (except for the critical value $\delta_c = -3^4 \mathcal{B}^2\mathcal{Q}^2 $). }

\begin{figure}[htb!]
\includegraphics[width=10 cm]{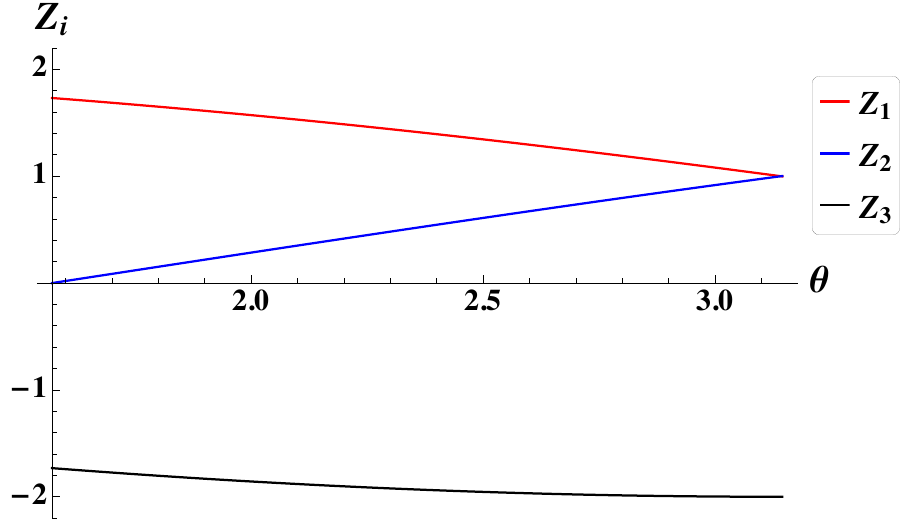}
\caption{The roots of the polynomial $H(z)$ considering the reparameterization $Z_i = z_i \sqrt{27 \mathcal{B}}$ .}
\label{fig3}
\end{figure}

If $\delta \in \mathbb{C}$, the set of $z_i$ renders three real roots, two of them positive and a third negative \textcolor{black}{as we may demonstrate in what follows. With the definition of the new constants}
\be
\alpha & \equiv & -81\mathcal{B}^2\mathcal{Q}^2 \\
\beta & \equiv & \sqrt{ \frac{4\mathcal{B}^3}{3}-\alpha^2} \\
\theta & \equiv & \arctan (\beta / \alpha )
\ee
\textcolor{black}{we can write $\delta = \alpha + i \beta $ and rewrite (\ref{z1}-\ref{z3}) as}
\be
\label{z1b}
z_1 & = & \frac{1}{\sqrt{27\mathcal{B}}}\Big(2\cos (\theta /3)  \Big) \\
\label{z2b}
z_2 & = & \frac{1}{\sqrt{27\mathcal{B}}}\Big(-\cos (\theta /3) +\sqrt{3}\sin (\theta /3) \Big) \\
\label{z3b}
z_3 & = & \frac{1}{\sqrt{27\mathcal{B}}}\Big(-\cos (\theta /3) -\sqrt{3}\sin (\theta /3) \Big) 
\ee
\textcolor{black}{Here it is important to notice that, since $\beta > 0$ and $\alpha < 0$, $\theta$ must be in $ \frac{\pi}{2} < \theta < \pi \equiv  \mathcal{R}_\theta $. Then the solutions (\ref{z1b}-\ref{z3b}) are constrained within $\mathcal{R}_c$ since $\tan \theta < 0$ in that. The equation satisfy very nicely the Girard relations for the solutions of a cubic polynomial as expected. In the range $\mathcal{R}_c$ we have that $z_1 > 0$ always and $-\cos (\theta /3) + \sqrt{3} \sin (\theta /3) > 0 $ as a consequence of $Max (\sqrt{3} \tan \theta /3 ) > 1 \hspace{0.1cm}( \theta \in \mathcal{R}_\theta ) $. This brings as a consequence that $z_2>0$ and $z_3$<0.} In figure \ref{fig3} we display all $z_i$ functions unraveling the critical point for the geometric parameters $\mathcal{B}$ and $\mathcal{Q}$: once $z_1 = z_2$ we have two horizons coalesced to a single $z_i$ defining the threshold for the geometry as $\theta = \pi$ and $\beta = 0$.
 We apprehend that the critical condition for the cosmological term as
\be
\label{crit2}
\mathcal{B}_{c}\mathcal{Q}_{c}^4 = B_{c}Q_{c}^4 = \frac{4}{3^9} \sim 0.000203221
\ee
representing the maximum value that $B$ can attain in the Chapligyn geometry, beyond which the spacetime presents only one horizon (cosmological). We display a quasi-extremal geometry for the lapse function in \ref{fig2}. 

In the analysis we demonstrate the existence of two critical points beyond which the spacetime presents only one horizon, either cosmological, $r_\mathcal{B}$ or, Cauchy $r_c$. In both cases the asymptotic region, $r> r_h$ is dynamically defined by a negative lapse function, similar to that of a Reissner-Nordström-de Sitter black hole when the cosmological supersedes a critical value. 

\textcolor{black}{Both of the thresholds (\ref{crit1}) and (\ref{crit2}) represent the critical values above which the geometry is covered by only one horizon (cosmological). They can not provide however the limits on $\mathcal{B}$ for a particular $\mathcal{Q}$ or vice-versa. The determination of each critical point relating $\mathcal{B}$ and $\mathcal{Q}$ is achieved with the following procedure: choosing a particular $\mathcal{Q}_p< \mathcal{Q}_{th}$, the equation 
\be
g\Big|_{\mathcal{Q}=\mathcal{Q}_p} = 0
\ee
provides exactly two solutions, $y_1$ and $y_2$. In turn, those two solutions of $g$ when substituted as singular points of the equation (\ref{lap3})
\be
f(y_1) = 0, \hspace{1.0cm} f(y_2) = 0
\ee
set the maximum and minimum values of $\mathcal{B}$. Such procedure can be performed only numerically and we provide a plot on the thresholds relating both properties for the existence of a three horizons black hole in figure \ref{figth}.
\begin{figure}[htb!]
\includegraphics[width=10 cm]{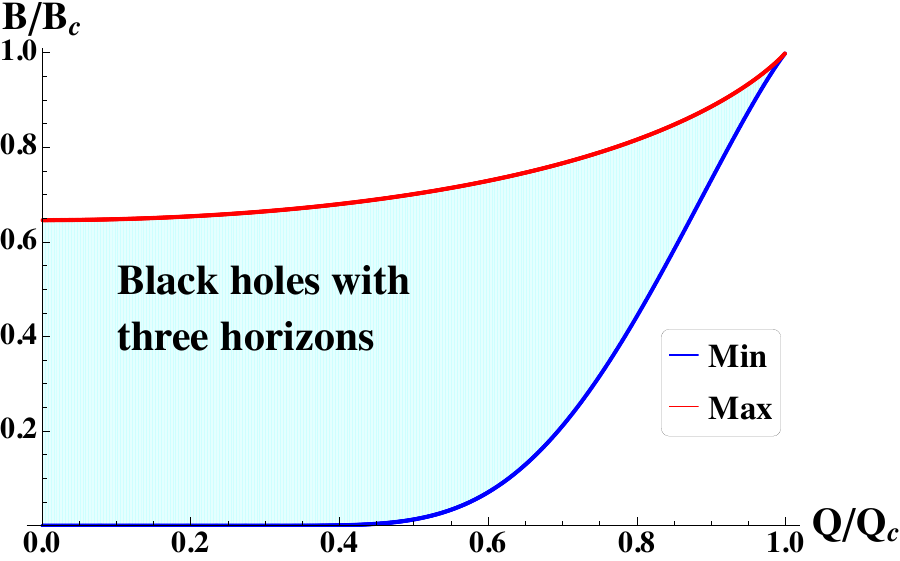}
\caption{\textcolor{black}{The threshold curves of a three horizon solution encoverin a Chaplygin singularity. Here $Q_{c}$ ist determined via (\ref{crit1}-\ref{crit1a}) as $Q_{c} = 2M \mathcal{Q}_{th}  $ and $B_{c} = \frac{4}{3^9Q_{c}^4}$ via (\ref{crit2}).}}
\label{figth}
\end{figure}
In Fig. \ref{figth}, we observe that for each value of $\mathcal{Q}$ there exists a maximum and a minimum value of $\mathcal{B}$ for which the black hole solution remains physically acceptable, exhibiting three horizons. In contrast to its Reissner–Nordström–de Sitter counterpart, the Chaplygin solution requires a minimum value of $\mathcal{B}$ (associated with the cosmological term) in order to remain well behaved.}

\textcolor{black}{
To conclude this section, we present the necessary formulae for a final panel showing a three horizon-Penrose diagram, which highlights the causally connected regions of the manifold endowed with the Chaplygin metric.
We begin by rewriting the lapse function $f(r)$ given in eq. (\ref{lapse}) in terms of its three real positive roots $(r_-, r_+, r_c)$, which correspond, respectively, to the cosmological horizon, the event horizon, and the Cauchy horizon with $r_- < r_+ < r_c$.
}
\textcolor{black}{\begin{equation}\label{lapse_factor}
f(r) \equiv F(r)G(r) \equiv \frac{\sqrt{B}}{r^2}\,(r-r_c)(r-r_+)(r-r_-)G(r).
\end{equation}}
\textcolor{black}{Since the singular points are restricted to $F$ and $G=0$ presents no real solution, the important part of the tortoise coordinate $r_{*}$ (that reveals its causal diagram) can defined as usual by
\begin{equation}
r_*=\int \frac{dr}{F(r)},
\end{equation}
where
\begin{equation}\label{lapso_frac}
\frac{1}{F} = -\frac{1}{k_-} \frac{1}{(r- r_-)} +\frac{1}{k_+}\frac{1}{(r-r_{+})} -\frac{1}{k_c} \frac{1}{(r-r_c)},    
\end{equation}
with $k_-$, $k_+$ and $k_c$ being the superficial gravity at Cauchy, event and cosmological horizons respectively. So, the tortoise coordinate $r_*$ reads as\footnote{The complete profile of the tortoise coordinate can be obtained only numerically as \begin{equation}
    r_{*} = -\frac{G_1(r)}{k_-}\log{|r-r_-|} +\frac{G_2(r)}{2k_+}\log{|r-r_+|} -\frac{G_3(r)}{k_c} \log{|r-r_c|}.
\end{equation}
in which $G_i (r)$ are infinite polynomials of $r$ with no real singular points. For our purposes, those functions acquire numerical values in each part of the diagram and can be reincorporated in the gravity sufaces, $k_i$.}
\begin{equation}
    r_{*} = -\frac{1}{k_-}\log{|r-r_-|} +\frac{1}{2k_+}\log{|r-r_+|} -\frac{1}{k_c} \log{|r-r_c|}.
\end{equation}}
\textcolor{black}{From the tortoise coordinate, we introduce the usual double-null coordinates
\begin{equation}
u=t-r_*,
\qquad
v=t+r_* .
\end{equation}}
\textcolor{black}{We now perform a analysis of the geometry in the neighborhood of each horizon in order to construct the Kruskal type extensions and the Penrose diagram.}

\textcolor{black}{Near the cosmological horizon $r\approx r_c$, the lapse function can be written as}
\textcolor{black}{\begin{equation}\label{lapse_cosmo}
f(r) = \mp 2k_c|r-r_c|,    
\end{equation}
where the the plus sign means $r>r_c$ and the minus sign to $r<r_c$, suggesting the following set of Kruskal-type coordinates near the cosmological horizon:}
\textcolor{black}{\begin{equation}
U_c=\pm e^{k_c u},
\qquad
V_c=e^{V/(2\alpha_c)},
\end{equation}
plus sign refers to $r>r_c$ and minus to $r<r_c$,  so}
\textcolor{black}{\begin{equation}\label{kruskal_cosmo}
U_c V_c
= \pm \frac{|r-r_c||r-r_-|^{k_c/k_-}}{|r-r_+|^{k_c/k_+}} .
\end{equation}}
\textcolor{black}{Following the same steps for neighborhood of the event horizon $r_+$ and Cauchy horizon $r_-$, we obtain the following sets of Kruskal-type coordinates,}

\begin{figure}[htb!]
\includegraphics[width=13 cm]{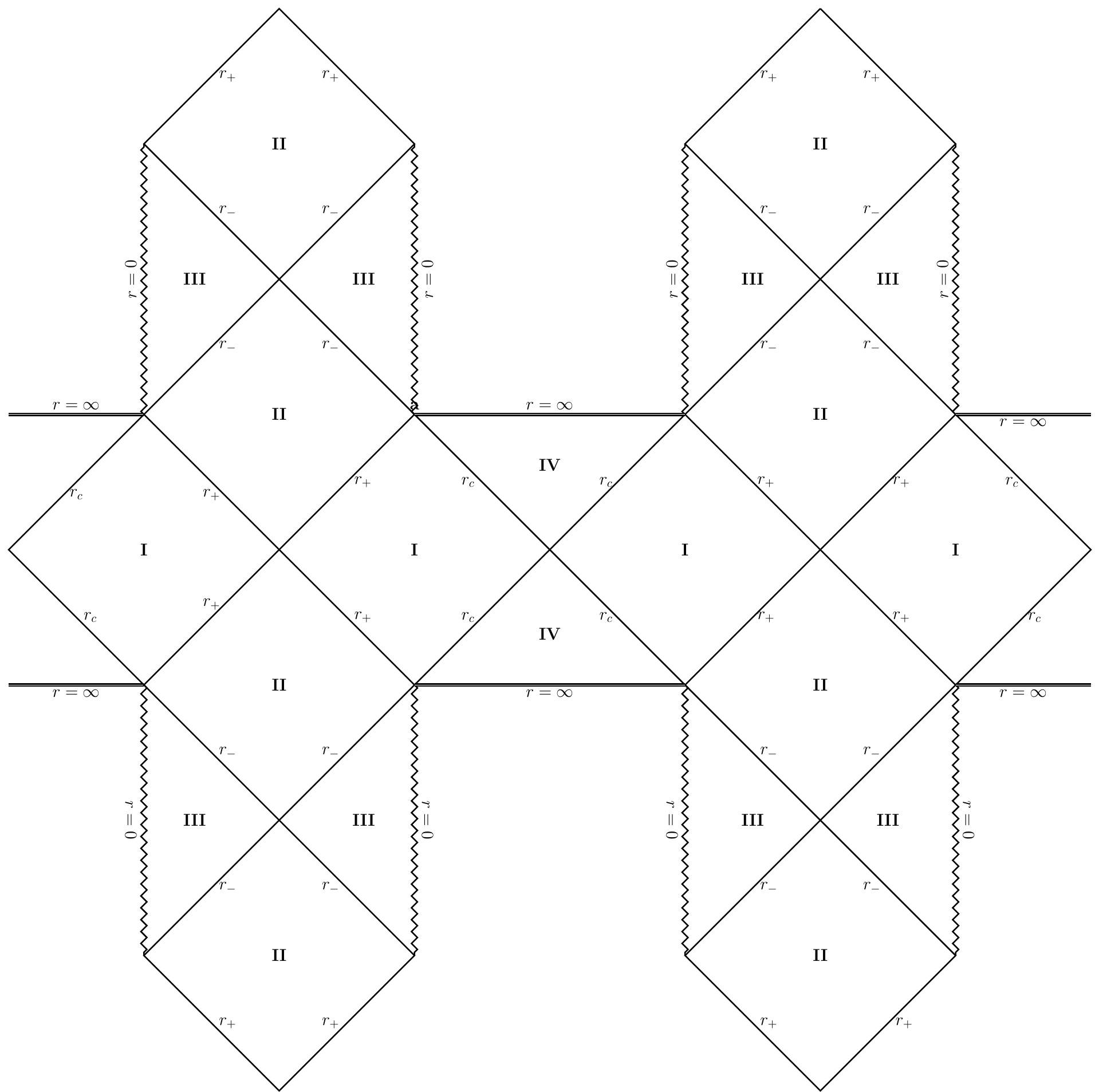}
\caption{\textcolor{black}{Penrose-Carter diagram for the black hole imersed in the Chaplygin dark fluid. The causal structre is identical to the Reissner-Nordstrom-de Sitter black hole. }}
\label{fig_diagrama}
\end{figure}

\textcolor{black}{\begin{equation}\label{kruskal_event}
U_+ V_+
= \mp \frac{|r-r_+|}{|r-r_-|^{k_+/k_-}|r-r_c|^{k_+/k_c}}, 
\end{equation}}
\textcolor{black}{\begin{equation}\label{kruskal_event}
U_- V_- = \pm \frac{|r-r_-||r-r_c|^{k_-/k_c}}{|r-r_+|^{k_-/k_+}}. 
\end{equation}}
\textcolor{black}{To construct the Penrose diagram we introduce double-null coordinates adapted to each root $(r_-,r_+,r_c)$ of the lapse function $f(r)$. The compactification is then performed through
\[
\tilde U=\arctan U, \qquad \tilde V=\arctan V ,
\]
from which the Penrose coordinates are defined as
\[
T=\frac{1}{2}(\tilde U+\tilde V), \qquad 
R=\frac{1}{2}(\tilde U-\tilde V).
\]
By matching the overlapping coordinate patches across the different horizons, the metric can be smoothly extended through them, allowing the construction of the conformal diagram. The resulting causal structure in Fig.  (\ref{fig_diagrama}) coincides with that of the Reissner-Nordström-de Sitter geometry, displaying an infinite sequence of regions separated by the Cauchy ($r_-$), event ($r_+$), and cosmological horizons ($r_c$), together with a timelike singularity located at $r=0$. Consequently, although the metric function is modified by the presence of the Chaplygin dark fluid, the global causal structure of the spacetime remains the same as that of the standard Reissner-Nordström-de Sitter solution. }


\section{Discussion and final remarks}\label{sec4}

In this work, we performed an analysis of the causal structure of spherically symmetric black holes immersed in a Chaplygin-like dark fluid, emphasizing how the nonlinear equation of state $p=-B/\rho$ modifies the spacetime geometry beyond standard vacuum and cosmological black hole solutions. By explicitly comparing our results with the Reissner–Nordstr\"om and Reissner–Nordstr\"om-de Sitter cases, we showed that the presence of the Chaplygin fluid enhances spacetime curvature, particularly in the region near the physical singularity at $r=0$.

One of the central results of this analysis concerns the threshold separating non-extremal black holes from their extremal configurations. We demonstrated that spherically symmetric black holes supported by a Chaplygin-like fluid reach the extremal regime for significantly smaller values of the parameter $Q$ when compared to both the Reissner–Nordstr\"om and Reissner–Nordstr\"om–de Sitter spacetimes. So, the Chaplygin-like dark fluid lowers the extremality bound, making the horizon structure more sensitive to environmental parameters. 

In addition to the bound on extremality parameter $Q$, we obtained an upper bound for the Chaplygin cosmological parameter $B$. By considering the roots of the polynomial governing the stationary points of the lapse function $f(r)$, we derived the upper bound for $B$, namely, $B_c Q_c^4 = 4/3^9$, above which the geometry no longer admits black hole horizons. For $B<B_c$, the spacetime may exhibit an event horizon, a Cauchy horizon and a cosmological horizon, depending on the value of $Q$, including extremal configurations where the event e Cauchy horizons coalesce. In the case $B>B_c$, the black hole horizon structure disappears and the spacetime is characterized by a single cosmological horizon. This result shows that the Chaplygin-like dark fluid not only lowers the extremality threshold but also imposes a upper limit on the effective cosmological constant $\Lambda_{eff} = 3\sqrt{B}$ compatible with black hole solutions.

\begin{acknowledgments}
This work was partially supported by CNPq (\textit{Conselho Nacional de Desenvolvimento Científico - Brazil}) under Grants Nos. $407868/2025-9$ and $405749/2023-6$, and FAPEMAT (\textit{Fundação de Amparo à Pesquisa do Estado de Mato Grosso - Brazil}) under Grant No. $0002203/2025$.
\end{acknowledgments}





\end{document}